\documentclass[aps,prb,reprint,groupedaddress]{revtex4-2}
\usepackage{graphicx}
\usepackage{bm}
\usepackage{xcolor}

\begin{document}

\newcommand{\be}{\begin{equation}}
\newcommand{\ee}[1]{\label{#1}\end{equation}}
\newcommand{\bem}{\begin{eqnarray}}
\newcommand{\eem}[1]{\label{#1}\end{eqnarray}}
\newcommand{\eq}[1]{Eq.~(\ref{#1})}
\newcommand{\Eq}[1]{Equation~(\ref{#1})}
\newcommand{\ua}{\uparrow}
\newcommand{\da}{\downarrow}
\newcommand{\g}{\dagger}
\newcommand{\rc}[1]{\textcolor{red}{#1}}


\title{The theory of planar ballistic  SNS junctions at $T=0$}



\author{Edouard  B. Sonin}
\email[]{sonin@cc.huji.ac.il}

\affiliation{Racah Institute of Physics, Hebrew University of Jerusalem, Givat Ram, Jerusalem 9190401, Israel}


\date{\today}

\begin{abstract}
The Letter presents the theory of planar ballistic SNS junctions at $T=0$ for any  normal  layer thickness $L$ taking into account phase gradients in superconducting leads. The  current-phase relation was derived  in the model of the steplike pairing potential  analytically and is exact  in the limit of large ratio of the Fermi energy to the superconducting gap. At small $L$ (short junction) the obtained  current-phase relation is essentially different from that in the previous  theory neglecting phase gradients.  It was confirmed by recent numerical calculations and was observed in  the experiment on short InAs nanowire Josephson junctions. The analysis resolves the problem with the charge conservation law in the steplike pairing potential model.
     \end{abstract}


\maketitle

\begin{figure}[t]
\centering
\includegraphics[width=0.25\textwidth]{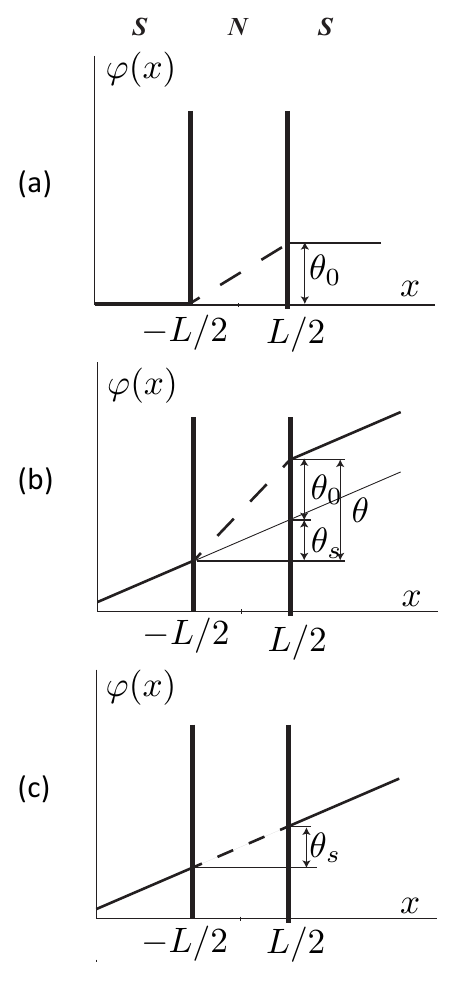}
\caption{The phase variation across the SNS junction. (a) The vacuum current produced by the vacuum phase $\theta_0$. The current is confined to the normal layer.   (b) The  superposition of the  vacuum current   and the condensate current determined by the superfluid phase $\theta_s =L\nabla \varphi$. The phase $\theta=\theta_0+\theta_s$ is the Josephson phase. (c) The condensate current produced by the phase gradient  $\nabla \varphi$ in the superconducting layers.  In all layers the electric current is equal to $env_s$.  }\label{f1}
\end{figure}

{\em Introduction}--The ballistic SNS junction has already been studied more than  a half-century. Starting from the pioneer  papers \cite{Kulik,Ishii,Bard} they used the self-consistent field method \cite{deGen}.  In this method  an effective pairing potential is introduced, which transforms the second-quantization Hamiltonian with the electron interaction into an effective Hamiltonian quadratic in creation and annihilation electron operators. The effective Hamiltonian   can be diagonalized by the Bogolyubov\,\textendash\,Valatin transformation. 

The effective Hamiltonian is not gauge invariant, and the theory using it  violates the charge conservation  law. The charge conservation law is restored  if one solves the Bogolyubov\,\textendash\,de Gennes equations  together with the self-consistency  equation for the pairing potential. 
     
The widely accepted existing theory \cite{And65,Kulik,Ishii,Bard,JJbk,ThunKink,Green}, which is the background of the present work, introduced some concepts and used some assumptions simplifying the analysis:

\begin{enumerate}
\item Starting from the original paper of \citet{And65}, instead of solving the self-consistency  equation,  it was assumed   that there is  a gap $\Delta$ of constant modulus $\Delta_0=|\Delta |$  in the superconducting layers and zero gap inside the normal layer. This means that the proximity effect is ignored. Further we call it the steplike pairing potential model. 

\item  The effective masses and Fermi velocities were assumed to be the same in  all layers. At this assumption, the normal scattering is absent, and  only the Andreev scattering is possible   at the SN interface.  

\item   Another outcome of the assumption of equal masses and Fermi velocities wa     was that the original second-order differential Bogolyubov\,\textendash\,de Gennes equations equations are reduced to the first order equations with boundary conditions imposed only on the wave function, but not on its derivative.

\item  It was assumed that not only the absolute value $\Delta_0$ but also phases were constant in superconducting leads (no phase gradients). Constant phases in two leads, however, were different, as shown in Fig.~\ref{f1}(a). 

\end{enumerate}

Under these assumptions the problem is reduced to quadratures. There is the {\em ab initio} exact expression for the current through the junction via the sum over all Andreev bound states and the integral over all continuum states. But these sum and integral are rather complicated for analytical and even numerical calculation because of oscillating integrand and necessity to calculate small difference of large terms. The rather sophisticated formalism  of temperature Green function was used \cite{Kulik,Ishii,ThunKink}.

At the phase profile  in Fig.~\ref{f1}(a) the charge conservation law is violated since the current flows only inside the normal layer. But it was believed that the charge conservation law can be restored by so small phase gradients in leads that this cannot affect calculations  ignoring the gradients. This suggestion is  true for a weak link, inside which the phase varies much faster than in the leads. 
 But planar SNS junctions are not weak links at zero temperature \cite{Son21}, and  phase gradients in the leads do affect the current in the normal layer. Thus, one should determine   currents in the normal and  superconducting layers  self-consistently.  

  In Refs.~\cite{Son21,SonAndr}   the assumption of constant phases in superconducting leads was revised for planar SNS junctions, in which  transverse cross-sections of all layers are the same. 
  We shall use the adjective ``planar''  even for 1D wires when the plane cross-section becomes a point. The motivation was an intention to restore the charge conservation law WITHIN the steplike pairing potential model instead of postponing this for more sophisticated calculations taking into  account  the self-consistency  equation.

       The following gap profile with the constant gradient $\nabla \varphi$ in leads was considered [Fig.~\ref{f1}(b)]:
\be
\Delta =\left\{ \begin{array}{cc} \Delta_0e^{i\theta_0+ i\nabla \varphi x} & x>L/2 \\  0 & -L/2<x<L/2\\ \Delta_0e^{i\nabla \varphi x} & x<-L/2  \end{array} \right. .
   \ee{prof}
The phase profile is determined by two phases $\theta_0$ and $\theta_s=L\nabla\varphi$. There is a mathematically correct  exact analytical  solution of the Bogolyubov\,\textendash\,de Gennes equations for any choice of $\theta_0$ and $\theta_s$. But we filter these solutions by the requirement of the charge conservation law. The strict conservation law was replaced by  a softer condition that, at least,  total currents deep in all layers are the same. 

The state with the phase gradients in leads [Fig.~\ref{f1}(b)] can be obtained from the state without gradients [Fig.~\ref{f1}(a)] by the Galilean transformation.  Galilean invariance of the ballistic SNS junction despite broken translational invariance has been already noticed by \citet{Bard} and confirmed in Ref.~\cite{Son21} in the steplike pairing potential model.  In Ref.~\cite{SonAndr} the Galilean invariance  was demonstrated for an arbitrary gap profile under the conditions that the ratio of $\Delta_0$ to the Fermi energy  is small and  the Andreev reflection is the only mechanism of scattering. 

The Galilean transformation of the ground state ($\theta_s=\theta_0=0$) yields the state with the phase profile shown in Fig.~\ref{f1}(c). In this state the total current $J$  in all layers is equal to the 
same current $J_s$ as in a uniform superconductor:
 \be
J=J_s=env_s=J_0{\theta_s\over \pi}=J_0{\theta\over \pi},
    \ee{Js}
where
\be
J_0={\pi en\hbar\over 2mL}={e\hbar k_f\over mL},
     \ee{} 
 $e$ is the electron charge, $m$ is the electron mass, $n$ is the electron density,  $v_s= {\hbar \over 2m}\nabla \varphi$ is the superfluid velocity,  and $k_f$ is the 1D Fermi wave number.  The phase $\theta_s$ and $J_s$ were called the superfluid phase and the condensate current respectively \cite{Son21}. 
 
 Since the Galilean invariance takes place for any gap variation in space,  the linear  current-phase relation (CPR) \eq{Js} is valid beyond  the steplike pairing potential model, and  the more accurate approach  taking into account the self-consistency equation cannot modify it. This agrees with  numerical calculations by \citet{Bagwell}.  They obtained that although the pairing potential amplitude  smoothly varied across  interfaces between layers, the phase gradient remained strictly constant along the whole junction as in Fig.~\ref{f1}(c).   Recent numerical calculations by \citet{Krekels} also confirmed this (see further discussion of this work in the end of the Letter). But nobody paid attention that these numerical calculations contradict the phase profile with constant phases in leads [Fig.~\ref{f1}(a)]   assumed in the existing theory.

The phase $\theta_0$ and the current $J_v$ produced by this phase were called the vacuum phase and the vacuum current respectively \cite{Son21}.The current  $J_v$ is calculated with all Andreev levels being unoccupied. Since  the current   $J_v$ flows only in the normal layer and violates the charge conservation law, it should be compensated by the current produced  by quasiparticles at Andreev levels. It was called excitation current $J_q$ \cite{Son21}. At zero temperature the excitation current appears at the critical current determined by the Landau criterion that the energy of the lowest Andreev level reaches 0. The CPR is derived from the condition that the total current flowing only in the normal layer vanishes: $J_v+J_q=0$. 

\begin{figure}[h]
\centering
\includegraphics[width=0.3\textwidth]{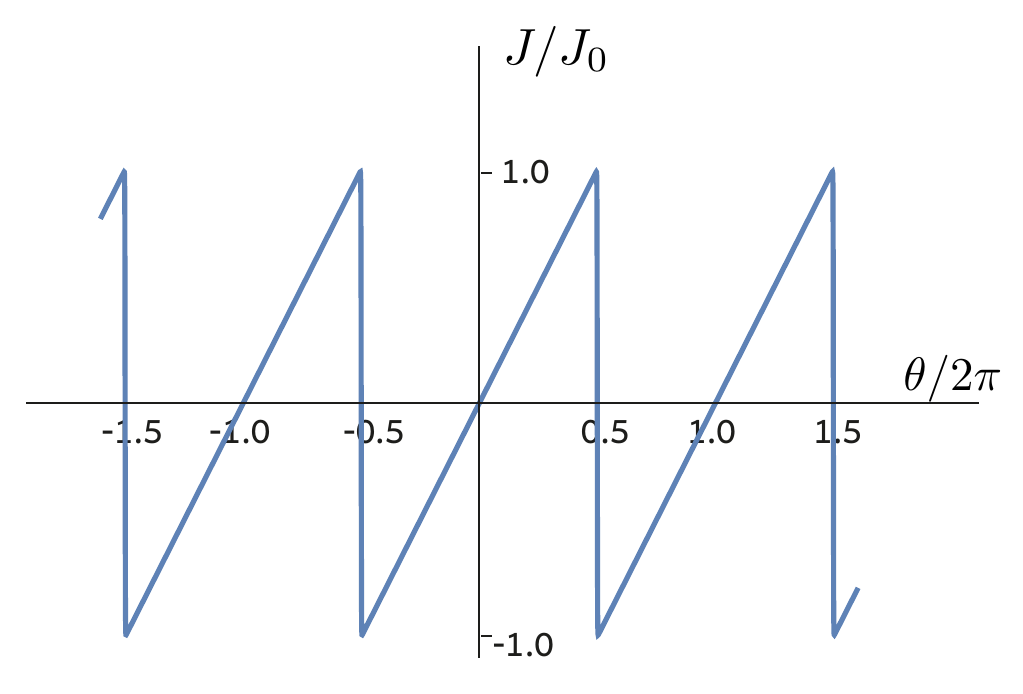}
\caption{Saw-tooth CPR ($T=0$, $L\to \infty$).}\label{ST}
\end{figure}

Summarizing, contrary to the existing theory, which connected the total current only with the vacuum current,  the condensate current can flow   through the planar SNS junction without violation of the charge conservation law at the same phase profile as in a uniform superconductor. However, despite the essential difference in the physical picture of the charge transport through the junction, at zero temperature the both theories predicted in the limit $L \to \infty$ (long junction) the same saw-tooth CPR (Fig.~\ref{ST}) given by \eq{Js}  at $-\pi <\theta<\pi$. 
Coincidence of the two CPRs  in the limit $L \to \infty$  led to the wrong conclusion that there is no difference between the condensate and the vacuum currents (see discussion in Refs.~\cite{Thun,Son23rep}). 

Up to now the analysis \cite{Son21,SonAndr} was focused on long junctions with $L$ much larger than the coherence length
\be
\zeta_0={\hbar^2 k_f\over m \Delta_0}. 
     \ee{calL}
The present Letter  addresses the whole diapason of $L$ down to $L = 0$.  Remarkably, at zero temperature it is possible to obtain a simple CPR analytically  for any $L$. The difference between the two theories  grows with decreasing $L$ and is maximal at $L=0$. In the limit $L=0$ the normal layer disappears, and the SNS junction becomes a uniform superconductor. The theory neglecting phase gradients  contradicts this evident assertion.

The analysis in the Letter mostly addresses the 1D case (single quantum channel). For the extension of the calculation  on multidimensional  (2D and 3D) systems
currents are integrated over  transverse wave vector components.

{\em Bogolyubov\,\textendash\,de Gennes equations, Galilean invariance,  and the energies of Andreev levels}--The  Bogolyubov\,\textendash\,de Gennes equations for the   wave function 
\begin{equation}
\psi(x) = \left[ \begin{array}{c} u(x)\\  v(x) \end{array}
\right]
     \label{spinor} \end{equation}
with the energy $\varepsilon$ are
\bem
\varepsilon  u  =-{\hbar^2 \over 2m} \left( \nabla^2 + k_f^2\right) u  + \Delta  v ,
\nonumber \\
\varepsilon v  \epsilon={\hbar^2 \over 2m}  \left( \nabla ^2 + k_f^2\right) v  + \Delta^* u .
     \eem{BG}

As in the previous investigations, it is assumed that the Fermi energy $\varepsilon_f={\hbar^2k_f^2\over 2m}$ is  much larger than the gap $\Delta_0 $. Then 
only Andreev reflection is possible, and there is no significant change of the quasiparticle momentum after reflection. 
Wave functions are superpositions  of plane  waves with wave numbers only close to either  $+ k_f$, or  $- k_f$. These plane waves describe quasiparticles, which  will be  called  rightmovers (+) and leftmovers (-).  After transformation of the wave function,
\be
\left(  \begin{array}{c}
u    \\
  v 
\end{array}  \right) =\left(  \begin{array}{c}
\tilde u    \\
\tilde  v 
\end{array}  \right)e^{\pm ik_f x},
        \ee{}
the second order  terms   in gradients, $\nabla^2 \tilde  u$ and $\nabla^2 \tilde  v$, can be neglected for small $\Delta_0/\varepsilon_f$, and the second order Bogolyubov\,\textendash\,de Gennes equations are reduced to the equations of the first order  in gradients  \cite{And65}:
\bem
\varepsilon \tilde u  =\mp  {i\hbar^2 k_f\over m} \nabla \tilde  u  + \Delta \tilde  v ,
\nonumber \\
\varepsilon \tilde v =\pm {i\hbar^2 k_f\over m} \nabla \tilde v  + \Delta^*\tilde  u .
    \eem{BG1}

Let us demonstrate Galilean invariance of the Bogolyubov\,\textendash\,de Gennes equations when the wave functions are superpositions of only rightmovers, or only of leftmovers \cite{SonAndr}. Suppose that we found the Bogolyubov\,\textendash\,de Gennes function    $\left(  \begin{array}{c}
\tilde u_0    \\
\tilde  v_0 
\end{array}  \right)$ with the energy $\varepsilon_0$
 for an arbitrary profile of the superconducting gap $\Delta(x)$. One can check that the  wave function 
 \be
 \left(  \begin{array}{c}
\tilde u    \\
\tilde  v
\end{array}  \right)=\left(  \begin{array}{c}
\tilde u_0 e^{i\nabla\varphi x/2}   \\
\tilde  v_0  e^{-i\nabla\varphi x/2} 
\end{array}  \right)
        \ee{} 
with constant gradient $\nabla\varphi$ satisfies \eq{BG1} where the gap $\Delta(x)$ is replaced by    $\Delta(x)e^{i\nabla\varphi x}$ and the energy  is 
\be
 \varepsilon=\varepsilon_0 \pm   {\hbar^2 k_f\over 2m}\nabla\varphi=\varepsilon_0 \pm   \hbar k_f v_s=\varepsilon_0 \pm    {\hbar^2 k_f\over 2mL}\theta_s .
      \ee{GT}   
Thus, the Galilean transformation produces the same Doppler shift in the energy as in a uniform superconductor. The Galilean transformation transforms the current $j_i$ in any $i$th state (either bound or continuum)  to the current $j_i+en_i v_s$, where $n_i$ is the density in the $i$th state.  Summation over all bound and continuum states  yields that the Galilean transformation added the same  condensate current $J_s=env_s$ in all layers (since the density $n$ is the same in all layers), which does not violate the charge conservation law.  Our derivation valid for any profile of the gap in the space and  remains valid if $\Delta$ vanishes in some part of the space.

In the absence of phase gradients the energies of Andreev levels are known from numerous previous publication. They are determined by the equation  (see, e.g., \citet{Bard}, or Eq.~(2) of \citet{ThunKink}):
\be
\varepsilon_{0\pm} (s,\theta_0)={\hbar^2 k_f\over 2mL}\left[2\pi  s+2 \arccos {\varepsilon_{0\pm} (s,\theta_0)\over \Delta_0} \pm \theta_0\right],
   \ee{eps00}
where $s$ is an integer. 
For a moving condensate the energy 
 \be
\varepsilon_\pm (s,\theta)={\hbar^2 k_f\over 2mL}\left[2\pi  s+2\arccos {\varepsilon_{0\pm} (s,\theta_0)\over \Delta_0} 
 \pm \theta\right],
   \ee{eps0}
differs from the energy $\varepsilon_{0\pm} (s,\theta_0)$ by the Doppler shift.

For our derivation of the CPR only the lowest Andreev level of leftmovers (electrons moving in the direction opposite to the current direction) with the energy $\varepsilon_{0-}(0,\theta_0)$ in the  frame moving with the condensate is needed:
\be
\varepsilon_0(\theta_0)={\hbar^2 k_f\over 2mL}\left[2\arccos {\varepsilon_{0} (\theta_0)\over \Delta_0} 
 - \theta_0\right].
   \ee{LA}
The energy of this level in the laboratory frame is
\be
\varepsilon(\theta)={\hbar^2 k_f\over 2mL}\left[2\arccos {\varepsilon_{0} (\theta_0)\over \Delta_0} 
 - \theta\right].
   \ee{LAf}
In the limit $L\to 0$, the $s=0$ Andreev level of leftmovers is the only Andreev level with the energy \cite{KulOmel2}
\be
\varepsilon_0(\theta_0)=\Delta_0\cos{\theta_0\over 2}.
  \ee{ASJ}

The current in the Andreev state is determined by the canonical relation connecting it with the derivative of the  energy with respect to the phase:
\be
j_\pm(s,\theta_0)={2e\over \hbar}{\partial  \varepsilon_{0\pm} (s,\theta_0)\over \partial \theta_0}.
       \ee{js}
 The factor 2 takes into account that $\theta_0$ is the phase of a Cooper pair but not of a single electron.

The current $j_\pm(s)$ is a current produced by a quasiparticle created at the $s$th state. In  unoccupied Andreev states the vacuum current  is two times less and has a  sign opposite to the sign of $j_\pm(s)$ \cite{SonAndr}. At $L\to 0$ the vacuum current in the Andreev state with the energy  \eq{ASJ} is 
 \be
 j_v(\theta_0)=-{e\over \hbar}{\partial  \varepsilon_0 (\theta_0)\over \partial \theta_0}= {e\Delta_0\over 2 \hbar}\sin{\theta_0\over 2}.
 \ee{jL}

{\em Current-phase relation}--At  Josephson phase $\theta$ smaller than the critical value $\theta_{cr}$  (see below) the condensate current is the only current in all layers, and   the CPR is given by \eq{Js}.   

The phase $\theta_{cr}$ is  determined from the Landau criterion: at $\theta=\theta_{cr}$  the energy of the lowest Andreev level $\varepsilon(\theta_0)$ in the laboratory coordinate frame reaches zero.  At $\theta>\theta_{cr}$  a nonzero vacuum current appears since $\theta_0 \neq 0$. The CPR must be derived from the condition $\varepsilon(\theta_0)=0$, which allows occupation of the  lowest Andreev level. 
At  $\varepsilon(\theta_0)=0$ Eqs.~(\ref{GT}) and   (\ref{LAf})  give relations
\be
\varepsilon_0(\theta_0) - {\hbar^2 k_f\over 2mL}\theta_s=0,~~\arccos {\varepsilon_0(\theta_0)\over \Delta_0} ={\theta\over 2}.
       \ee{}
Together with \eq{Js} connecting the total current $J$ with  $\theta_s$, this yields the exact CPR 
\be 
J= J_{cr}  \cos {\theta\over 2},
     \ee{PSJ}
where     
\be 
J_{cr}= {2 e\Delta_0 \over \pi \hbar} 
     \ee{Jcr}
is the critical current in the superconducting leads determined from the Landau criterion (depairing current).

The critical phase $\theta_{cr}$ is determined by the condition that the two CPRs [Eqs.~(\ref{Js}) and (\ref{PSJ})] give the same current:
\be
\theta_{cr}- {2m L\Delta_0 \over \hbar^2 k_f}\cos {\theta_{cr}\over 2}=\theta_{cr}- {2L \over \zeta_0}\cos {\theta_{cr}\over 2}=0.  
     \ee{thCr}

\begin{figure}[h]
\centering
\includegraphics[width=0.3\textwidth]{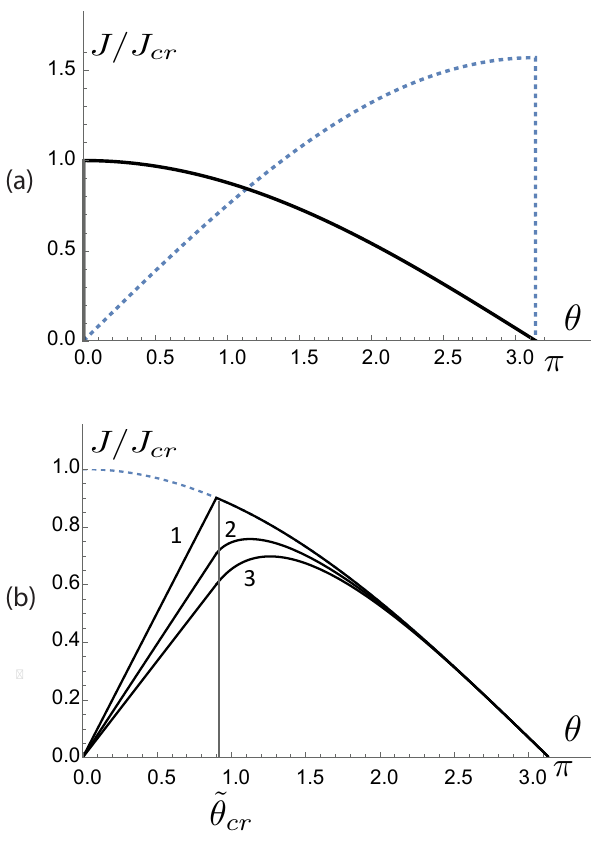}
\caption{Current-phase relations at $T=0$. (a) $L=0$.   The  solid  line shows the current phase relation valid for any dimensionality of the junction. The current phase relation in the theory neglecting phase gradients in leads  is shown by the dashed line.
(b) $L=\tilde \zeta/2$. The curves 1, 2, and 3 are the current phase relations for  1D, 2D, and 3D junctions respectively. In the 1D case the length  $\tilde \zeta$ coincides with   $ \zeta_0$. \label{ShJ}}
\end{figure}

 The part of the CPR at $\theta <\theta_{cr}$ can be called   condensate current branch. In this branch the condensate current is the only current through the junction ($\theta_0=0$, $\theta=\theta_s$), and the phase distribution is the same as in a uniform superconductor [Fig.~\ref{f1}(c)]. At $\theta>\theta_{cr} $  the vacuum current and the excitation currents appear ($\theta_0\neq 0$ and $\theta_s\neq \theta$), but their sum vanishes, as required by the charge conservation law. Along this branch the phase slip occurs when the phase difference across the junction loses  $2\pi$. So, the branch can be called phase slip branch. The CPRs for 1D junctions are shown for $L=0$ in Fig.~\ref{ShJ}(a) (solid line) and for $L/\zeta_0=1/2$  in Fig.~\ref{ShJ}(b) (curve 1). 

The  condensate current  branch at zero phase in Fig.~\ref{ShJ}(a) is vertical because at  $L=0$ the junction becomes a uniform superconductor, in which there are no phase jumps at currents lees than critical. The phase branch at $L= 0$  describes a phase slip in a uniform superconductor. 

When the normal layer thickness $L$ grows the slope of the condensate current branch  and the Josephson critical current  (the maximal current across the junction) decrease, while the critical phase $\theta_{cr}$ approaches  $\pi$. Eventually, the CPR is described by the saw-tooth current-phase curve (Fig.~\ref{ST}) for a very long junction with the small Josephson critical current   $J_0\propto 1/L$.

For derivation of the CPR  it was not necessary to make complicated calculations of sums and integrals determining the vacuum current.
Whatever its value is, at $T=0$ the vacuum current  in the phase slip branch is compensated  by the excitation current in the partially occupied lowest Andreev level. The values of these  currents separately are  necessary only if one wants to know how  the occupation number of this level varies along  the phase slip branch.

 For multidimensional systems currents calculated for a single 1D channel  must be integrated over  wave vectors ${\bf k}_\perp$ transverse to the current direction keeping in mind that $k_f=\sqrt{k_F^2-k_\perp^2}$. Here $k_F$ is the Fermi radius of a multidimensional system. The integration operation is $\int_{-k_F}^{k_F}  {dk_\perp\over 2\pi}… $ in the 2D  case and $\int_0^{k_F}  {k_\perp\,dk_\perp\over 2\pi}… $ in the 3D case. After integration currents become current densities.

At the condensate current branch the linear CPR remains valid after integration, but one should replace the 1D density $n$ by the 2D or 3D density. The condensate current branch extends up the phase $\tilde \theta_{cr}$ determined by the equation
\be
\tilde\theta_{cr}- {2L \over\tilde \zeta}\cos {\tilde\theta_{cr}\over 2}=0,  
     \ee{}
which is similar  to \eq{thCr} for 1D junctions with $k_f$ replaced by $k_F$ and the 1D coherence length $\zeta_0$ [\eq{calL}] replaced by the   coherence length
\be
\tilde \zeta =  { \hbar^2 k_F \over  m \Delta_0} 
       \ee{}
 for multidimensional junctions. At $\theta =\tilde\theta_{cr}$ the transition to the phase slip branch occurs only at the maximal $k_f=k_F$. In all other channels with $k_f< k_F$,  the condensate current branch extends up to phases larger than $\tilde \theta_{cr}$. Thus,  at  $\theta >\tilde \theta_{cr}$ we have a mixture of channels with the condensate current branch at $k_f<k_c$ and  with the phase slip branch at $k_f>k_c$. Here 
\be
k_c= {2L\cos {\theta\over 2} \over \tilde \zeta\theta }k_F.
 \ee{}
 Finally, integration over all channels yields 
\begin{widetext}
\bem
J={2e\Delta_0 \over \pi^2 \hbar}\cos {\theta\over2}\int_0^{\sqrt{k_F^2-k_c^2}} dk+{e\hbar \over \pi^2 mL}\theta \int_{\sqrt{k_F^2-k_c^2}}^{k_F}\sqrt{k_F^2-k^2} dk
\nonumber \\
=J_{cr}\theta \left({\cos {\theta\over2}\over 2\theta}\sqrt{1-{4L^2\cos^2 {\theta\over 2} \over \tilde \zeta^2\theta^2 }}
+{\tilde \zeta \over 4L}\arctan{{2L\cos {\theta\over 2} \over \tilde \zeta \theta}\over \sqrt{1-{4L^2\cos^2 {\theta\over 2} \over \tilde \zeta^2 \theta^2}}}\right)
      \eem{2D}   
for the 2D junction and
\bem
J={e\Delta_0 \over \pi^2\hbar }\cos {\theta\over2} \int_0^{\sqrt{k_F^2-k_c^2}} k \,dk+{e\hbar \over 2\pi^2mL}\theta \int_{\sqrt{k_F^2-k_c^2}}^{k_F}\sqrt{k_F^2-k^2}2\pi k \,dk
=J_{cr}\cos {\theta\over2}\left(1-{4L^2\cos^2 {\theta\over 2} \over 3 \tilde \zeta^2\theta^2 }\right)
      \eem{3D}   
\end{widetext}
for the 3D junction. These expressions show that in the limit $L=0$ the expression for the ratio $J/J_{cr}$ in multidimensional junctions does not differ from that in 1D junctions, and the  plot $J/J_{cr}$ vs. $\theta$ [solid line in \ref{ShJ}(a)] describes the CPR for junctions of any dimensionality.  But  the critical current given by \eq{Jcr} for 1D junctions must be replaced by the critical current densities for multidimensional junctions:
\be
J_{cr}=\left\{ \begin{array}{cc}  {2 e\Delta_0 k_F \over \pi^2 \hbar}  &~~ 2D~\mbox{case} \\ \\  { e\Delta_0 k_F^2 \over 2\pi^2 \hbar} &~~  3D~\mbox{case} \end{array}\right. .
    \ee{}

The CPRs for 2D and 3D junctions at $L/\tilde \zeta=1/2$ are shown  in Fig.~\ref{ShJ}(b)  (curves 2 and 3) together with the CPR for a 1D junction (curve 1).  There is a cusp in the 1D CPR in the critical point $\theta=\tilde\theta_{cr}$, which is smeared out in the 2D and 3D cases.  In 2D and 3D junctions the first derivative (slope) of the current-phase curve is continuous at $\theta=\tilde\theta_{cr}$, but the critical point is still non-analytic with jumps in a higher derivative.

{\em Conclusions and discussion}--The Letter demonstrates that at $T=0$ phase gradients in the superconducting leads commonly  ignored in the past are important for planar ballistic SNS junctions for any thickness $L$  of the normal layer.  The  CPR was derived analytically and is exact   in the steplike pairing potential model for the Fermi energy much larger than the gap   $\Delta_0$.  A more realistic (but much more complicated)   theory taking into account the self-consistency equation can quantitatively  modify the phase slip branch of the CPR at the Josephson phase more than the critical value determined from the Landau criterion. But the condensate current branch at the Josephson phase less than critical value was obtained from the Galilean invariance and is valid beyond the steplike pairing potential model. 

The theory taking into account phase gradients in the superconducting leads resolves the problem of charge conservation law within the steplike pairing potential model. This refutes the opinion \cite{ThunKink} that the charge conservation law cannot be restored without solving the self-consistency equation. 

The difference between the previous theory and ours is especially large  for short junctions. One can see this in Fig.~\ref{ShJ}(a) showing the CPR for $L=0$. The solid line shows the CPR obtained in the present work. The dashed line shows the CPR obtained ignoring gradients in superconducting leads, i.e., assuming that $\theta_s=0$  and $\theta = \theta_0$. Within this approach,  the total current  at $L=0$ is 
\be
J=2j_v(\theta) = {e\Delta_0\over   \hbar}\sin{\theta\over 2},
    \ee{CFT} 
where  $j_v(\theta)$  is the vacuum current  in the single Andreev level [\eq{jL}].   The factor 2 takes into account two spins. 

\Eq{CFT}  was originally  derived by \citet{KulOmel2} for a superconducting bridge between bulk leads. It was widely accepted and discussed  in various papers and reviews (see, e.g., Eq.~(7) in Ref.~\cite{ThunKink}, Eq.~(177) in Ref.~\cite{Green}, or Eq.~(3.3) in Ref.~\cite{JJbk}). The present analysis demonstrates that  the CPR \eq{CFT} is invalid for planar SNS junctions  at $L\to 0$ \cite{SonCom}, when the  junction becomes a uniform superconductor. 

According to \eq{CFT}, the critical Josephson current is always at the phase $\pi$, i.e., the CPR is forward-skewed (the current maximum is located at the phase exceeding the phase  $\pi/2$ of the current maximum of the sinusoidal CPR). But at derivation of \eq{CFT} the charge conservation law was violated. According to our analysis repairing this flaw, in short junctions the CPR is backward-skewed (the current maximum is at the phase less than  $\pi/2$). 

In the past the possibility of backward-skewed CPR  was revealed in the theory of dirty SNS junctions at temperatures close to critical (see Sec.~2.10 in  \cite{Ivan} and Sec.~V.A.4 in \cite{GKI}). It was explained by  gap suppression close to the normal layer determined from the Ginzburg--Landau theory. This resulted in a nonlinear phase shift close to the junction. Far from the junction the phase was constant. In contrast, our work  revealed the backward-skewed CPR at $T=0$ in ballistic junctions at constant gap. In this case the charge conservation  requires constant phase gradients in superconducting leads.

\citet{InAs}  observed  the backward-skewed CPR  in short InAs nanowire junctions. They explained this by the effect of Coulomb interaction. According to Refs.~\cite{Ivan,GKI} and this Letter, the backward-skewed CPR is possible also in models without interaction.  But our theory is not directly applicable to their experiment on 1D normal  bridges between 3D leads, which are not planar junctions.

 \citet{Krekels} compared our theory with their  numerical calculations using a more accurate theory taking into account the self-consistency equation. They concluded that numerical results  qualitatively well  agree
with the backward-skewed CPR obtained in our analysis (cf.  Fig. 3(b) in this Letter and Fig. 3(a) in \cite{Krekels}) and disagree with the theory ignoring phase gradients in leads.

 The steplike pairing potential model is a strongly idealized model. Still the continuous interest to it  during 60 years is justified. As it was noted by  \citet{Bard} about  this model,  ``it is important to understand  physically the nature of the supercurrent flow for such an ideal problem without complications".
  
\begin{acknowledgments}
	I thank Vadim Geshkenbein for fruitful discussions.
\end{acknowledgments}


%

\end{document}